\documentclass[12pt]{amsart}
\usepackage{amssymb}
\input epsf

\theoremstyle{plain}
\newtheorem{theorem}{Theorem}[section]
\newtheorem{lemma}[theorem]{Lemma}
\newtheorem{proposition}[theorem]{Proposition}
\newtheorem{corollary}[theorem]{Corollary}
\newtheorem{algorithm}[theorem]{Algorithm}
\newtheorem{claim}[theorem]{Claim}

\theoremstyle{definition}
\newtheorem{definition}[theorem]{Definition}

\newtheorem{example}[theorem]{Example}

\theoremstyle{remark}
\newtheorem{remark}[theorem]{Remark}

\renewcommand{\(}{\left(}
\renewcommand{\)}{\right)}
\newcommand{\<}{\left<}
\renewcommand{\>}{\right>}
\newcommand{\note}[1]{\par\noindent{\sf\large Uzi, note:} {\sf #1}\par}
\long\def\forget#1\forgotten{}

\newcommand{\bits}{{{\mathbb F}_2}}
\newcommand{\x}{\times}
\newcommand{\lam}{\lambda}
\newcommand{\pol}[1]{{#1(\lam)}}
\newcommand{\s}{\sigma}
\newcommand{\p}{{\mathfrak p}}
\newcommand{\pq}{\p_{\pol{q}}}
\newcommand{\sbst}{\subseteq}

\begin{document}

%







\title[Word-oriented LFSR's]{Efficient linear feedback shift registers with maximal period}

\author{Boaz Tsaban}
\address{Department of Mathematics, Bar-Ilan University, 52900
Ramat-Gan, Israel}
\email{tsaban@macs.biu.ac.il}
\urladdr{http://www.cs.biu.ac.il/\~{}tsaban}
\author{Uzi Vishne}
\address{Landau research center for mathematical analysis,
Hebrew University of Jerusalem, Israel}
\email{vishne@math.huji.ac.il}

\begin{abstract}
We introduce and analyze an efficient family of linear
feedback shift registers (LFSR's) with maximal period.
This family is word-oriented and is suitable for implementation
in software, thus provides a solution to a recent challenge \cite{FSE94}.
The classical theory of LFSR's is extended to
provide efficient algorithms for generation of irreducible and primitive
LFSR's of this new type.
\end{abstract}

\keywords{linear feedback shift registers, linear transformation shift registers, fast software encryption}
\subjclass{11T06, 11T71}

\maketitle

\section{Linear feedback shift registers}

Linear feedback shift registers (\emph{LFSR}'s) are fundamental primitives in the theory and practice
of pseudorandom number generation and coding theory
(see, e.g., \cite{CAM}, \cite{GOL}, \cite{GON}, \cite{KAM}, \cite{MUN}, \cite{NOR}, and references therein).

\begin{figure}[!h]\label{lfsr}
$$\epsfysize=0.7 cm {\epsfbox{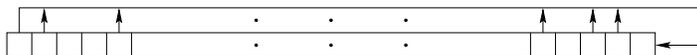}}$$
\caption{A typical LFSR.}
\end{figure}

Figure \ref{lfsr} describes a typical LFSR over the two-element field $\bits = \{0,1\}$,
where each step consists of adding some of the state bits
(we follow the convention that the elements of $\bits$ are called \emph{bits}),
and the result is inserted to the register in a FIFO manner.

Such a construction is slow in the sense that it produces only one new bit
per step.
Moreover, it is difficult to implement in software,
since many bit manipulations are required.
In certain cases (but not always \cite{VISH}),
it is possible to use LFSR's with only
two feedback taps. This makes a slightly faster LFSR.
(See also Section \ref{conclusions}.)

In the 1994 conference on fast software encryption, a challenge was set forth
to design LFSR's which exploit the parallelism offered by the word oriented operations
of modern processors \cite[\S 2.2]{FSE94}.
In this paper we suggest a solution and study its properties.

\section{Linear transformation shift registers}

Fix an arbitrary finite field $F$.
A sequence $\sigma = \<s_n\>_{n=0}^\infty$ of elements from $F$ is
\emph{linear recurring with characteristic polynomial}
$$\pol{f} = a_0 + a_1\lam + \dots + a_d\lam^d \in F[x]$$
if $a_d = 1$, and
$$a_0 s_n + a_1 s_{n+1} + \dots + a_d s_{n+d} = 0$$
for all $n=0,1,2,\dots$.
The \emph{minimal polynomial} of a linear recurring sequence $\sigma$ is
the characteristic polynomial of $\sigma$ of least degree.
Let $\sigma$ be a nonzero linear recurring sequence with an irreducible characteristic
polynomial $\pol{f}$. It is well known (cf.\ \cite{GOL}) that the period of $\sigma$
is equal to the order of $\lam$ in the multiplicative group of the field $K = F[\lam ]/\<\pol{f}\>$.
If $\lam$ generates the whole group, we say that $\pol{f}$ is \emph{primitive}.
(In this case $\sigma$ has the maximal possible period $|K|-1 = |F|^d-1$ where $d=\deg\pol{f}$.)
Likewise, for any natural number $d$, if $T$ is a linear transformation of $F^d$ and $v\in F^d$ is nonzero, then
the sequence $\<T^n(v)\>_{n=0}^\infty$ of vectors in $F^d$ has period $|F|^d-1$ if and only if the
characteristic polynomial of the linear transformation $T$ is
primitive over $F[\lam ]$. If this is the case we say that $T$ is \emph{primitive}.

We now introduce the family of \emph{linear transformation shift registers} (TSR's).
For convenience of presentation, we pack $m\cdot n$-dimensional vectors in an array $(v_0,\dots,v_{n-1})$ of
$n$ vectors in $F^m$ ($n$ and $m$ will be fixed throughout the paper).
In the intended application,
$F=\bits$ and $m$ is the number of bits in the processor's word. Typical values of $m$ are $8$, $16$, $24$, $32$, and $64$.
This way, the array $(v_0,\dots,v_{n-1})$ is stored in $n$ processor words.
Following this interpretation, elements of $F^m$ will be called \emph{words}.

\begin{definition}
Let $T$ be a linear transformation of $F^m$, and let $S=\<a_0,\dots,a_{n-1}\>\in F^n$.
A \emph{TSR step $\<T,S\>$} of the array $R=(v_0,\dots,v_{n-1})\in M_{m\x n}(F)$ is
the linear transformation
$$\<T,S\>(R) := (v_1,v_2,\dots,v_{n-1},T(a_0 v_0 + a_1 v_1 + \dots + a_{n-1} v_{n-1})).$$
The system $\<T,S,R\>$ is called a \emph{TSR}.
\end{definition}

\begin{figure}[!h]\label{TSR}
$$\epsfysize=4 truecm {\epsfbox{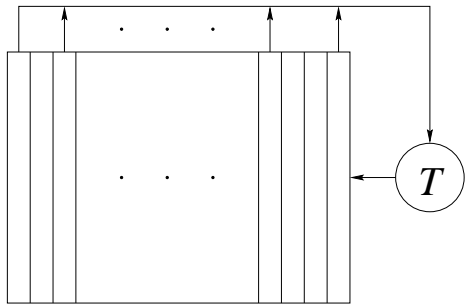}}$$
\caption{A typical TSR.}
\end{figure}

Figure \ref{TSR} illustrates a typical example of a TSR.
An obvious advantage over the standard LFSR is that here a whole new word (rather than a single bit)
is produced per step.

Linear transformations on processor words can
be performed very efficiently, either using lookup tables, or by using specific
linear transformations which are efficient when working on processor words, e.g.
Galois-type shift registers. The latter example has the
advantage that no additional memory is required (see, e.g., \cite[pp.~378--379]{SCH}).
Note further that choosing each of the $a_i$'s to be either $0$ or $1$ eliminates
the complexity of the multiplications $a_iv_i$.
One cannot, however, eliminate the complexity of the transformation $T$ as well by using
the identity transformation $T=I$: In this case the period cannot be greater than
$|F|^n-1$, whereas in principle, memory of $n$ words can yield period $|F|^{mn}-1$.

Simulations show that there exist choices for $T$ and $S$ such that the resulted TSR
step is primitive, and thus yields a sequence of vectors with period $2^{mn}-1$.
In the following sections we
provide necessary conditions on $T$ and $S$ in order that the resulted
TSR step is primitive.
Choosing $T$ and $S$ to satisfy these conditions increases the probability that the resulted TSR is primitive
with respect to random choice of these parameters.
Thus, we will get an efficient algorithm for generation of primitive TSR's.

\forget
Checking that a given TSR is primitive involves two tests. Only after one knows
that the characteristic polynomial of the TSR is irreducible, one proceeds to the
second test. The most time-consuming part is the search for a TSR which passes
the irreducibility test.
The rest of this paper is devoted to an analysis of the TSR transformations.
In particular, we will get an efficient algorithm for choosing $T$ and $S$
which pass the irreducibility test.
\forgotten

\section{The characteristic polynomial of a TSR}

Identify the linear transformation $T$ operating on words
with the matrix $T \in M_m(F)$ such that $T\cdot v = T(v)$, $v\in F^m$.

Let $I$ denote the $m\x m$ unit matrix.
A TSR step $\<T,S=\<a_0,\dots,a_{n-1}\>\>$ of the array $R=(v_0,\dots,v_{n-1})\in (F^m)^n$
is equivalent to multiplication of
$(v_0,\dots,v_{n-1})^t$ from the left
by the block matrix $[\<T, S\>] \in M_{nm}(F)$, where
$$
[\<T,S\>] = \(
\begin{array}{ccccc}
 0        & I       & 0         & \cdots    & 0    \\
 \vdots &\ddots & \ddots    & \ddots      & \vdots  \\
 \vdots &          & \ddots  & \ddots      & 0     \\
 0    &\cdots  & \cdots  & 0            & I    \\
 a_0T     & a_1 T   & \cdots    & a_{n-2}T  & a_{n-1}T
\end{array}
\).
$$

Let $\pol{f_S} = a_0 + a_1\lam + \dots+ a_{n-1}\lam^{n-1}$
(so that the characteristic polynomial of $[\<T,S\>]$ in the case $m=1$ and $T = (1)$ is $\lam^n-\pol{f_S}$),
and let $\pol{f_T} = |\lam I-T|$ denote the characteristic polynomial of $T$ (note that the degree of
$\pol{f_T}$ is $m$.)

\begin{proposition}\label{charpoly}
Let $T$ be a linear transformation of $F^m$, and $S = \<a_0,\dots,a_{n-1}\>\in F^n$.
Then the characteristic polynomial of the TSR step $\<T, S\>$ is
$$\pol{f_{\<T,S\>}} = \pol{f_S}^m \cdot f_T\(\frac{\lam^n}{\pol{f_S}}\).$$
\end{proposition}
\begin{proof}
We multiply each row block by $\lam$, and add the result to the next one.
Then we use the $-I$ blocks to cancel the terms in the first column block.

\begin{eqnarray}
|\lam I - \<T,S\> | & = & \left|
\begin{array}{ccccc}
 \lam I & -I & 0 & \cdots & 0 \\
 0  & \ddots & \ddots   & \ddots & 0 \\
 \vdots & \ddots& \ddots    & \ddots & 0    \\
 0  & \cdots & 0 & \lam I & -I \\
 -a_0T  & -a_1T & \cdots    & -a_{n-2}T & \lam I - a_{n-1}T \\
\end{array} \right| = \nonumber \\
& = & \left |
\begin{array}{ccccc}
 \lam I & -I & 0 & \cdots & 0 \\
 \lam ^2 I  & 0 & \ddots    & \ddots & 0 \\
 \vdots & \vdots & \ddots   & \ddots & 0    \\
 \lam ^{n-1}I   & 0 & \cdots & 0 & -I \\
 -a_0T  & -a_1T & \cdots    & -a_{n-2}T & \lam I - a_{n-1}T \\
\end{array} \right| = \nonumber \\
& = & \left |
\begin{array}{ccccc}
 0 & -I & 0 & \cdots & 0 \\
 \vdots & \ddots & \ddots   & \ddots & \vdots \\
 \vdots &  & \ddots & \ddots & 0    \\
 0 & \cdots & \cdots & 0 & -I \\
\lam^nI-\pol{f_S}T  & -a_1T & \cdots    & -a_{n-2}T & \lam I - a_{n-1}T \\
\end{array} \right| = \nonumber \\
& = & (-1)^{m(n-1)}\left |
\begin{array}{ccccc}
\lam^nI-\pol{f_S}T  & -a_1T & \cdots    & -a_{n-2}T & \lam I - a_{n-1}T \\
 0 & -I & 0 & \cdots & 0 \\
 \vdots & \ddots & \ddots   & \ddots & 0 \\
 \vdots &  & \ddots & \ddots & 0    \\
 0 & \cdots & \cdots & 0 & -I \\
\end{array} \right| = \nonumber \\
& = & (-1)^{m(n-1)}\cdot|\lam^n I -\pol{f_S}T|\cdot|-I|^{n-1} = \nonumber \\
& = & \pol{f_S}^m\cdot \left|\frac{\lam^n}{\pol{f_S}}I-T\right|
= \pol{f_S}^m\cdot f_T\(\frac{\lam^n}{\pol{f_S}}\). \nonumber
\end{eqnarray}
\end{proof}

A naive algorithm for generation of a TSR with maximal period would be to
choose the linear transformation $T$ and the set $S$ at random,
calculate the characteristic polynomial $\pol{f_{\<T,S\>}}$ using Proposition
\ref{charpoly},
and then check whether it is primitive, repeating this process until a
primitive polynomial is found.
In most of the cases, the polynomial will not be primitive for the
reason that it is not even irreducible.
The following corollary shows that much unnecessary work can be avoided.

\begin{corollary}\label{easyfactor}
If $\pol{f_T}$ is reducible over $F$, then so is $\pol{f_{\<T,S\>}}$.
\end{corollary}
\begin{proof}
Suppose $\pol{f_T} = \pol{q_1}\pol{q_2}$ is a nontrivial factorization of $\pol{f_T}$ over $F$, $m_i = \deg\pol{q_i}$. Then
$\pol{f_S}^{m_i}q_i\(\frac{\lam ^n}{\pol{f_S}}\)$ are
polynomials, and
$\pol{f_{\<T,S\>}} = \(\pol{f_S}^{m_1}q_1\(\frac{{\lam ^n}}{\pol{f_S}}\)\)\cdot\(\pol{f_S}^{m_2}q_1\(\frac{{\lam ^n}}{\pol{f_S}}\)\)$ is a
nontrivial factorization.
\end{proof}

\begin{remark}\label{speedup1}
In general, the probability that a monic polynomial
of degree $m$ chosen at random is irreducible is close to $1/m$.
Thus, by Corollary \ref{easyfactor}, the probability that $\pol{f_{\<T,S\>}}$ is irreducible provided that $\pol{f_T}$ is irreducible
should be about $m$ times larger than the probability when $\pol{f_T}$ is arbitrary.
\end{remark}

\section{Irreducibility through extension fields}

The algorithm stated in the previous section considered polynomials of a
special form as candidates to be primitive. In this section
we study polynomials of this form, with the aim of improving the algorithm.

Let $F$ be a fixed finite field.
Let $\pol{q} = q_0 + q_1\lam + \dots + q_m\lam^m \in F[\lam]$.
We write $\pq(x,y)$ for the homogeneous polynomial
$$x^m\cdot{q(y/x)} = q_0 x^m + q_1 x^{m-1} y + \dots + q_m y^m.$$
We wish to find necessary conditions for polynomials of the form
$\pq(\pol{g},\pol{f})$ to be irreducible.
Clearly, if $\pol{g},\pol{f}\in F[\lam]$ are not relatively prime, then
the polynomial $\pq(\pol{g},\pol{f})$ is reducible.
Also, by Corollary \ref{easyfactor}, if $\pol{q}$ is reducible, then so is
$\pq(\pol{g},\pol{f})$. We are thus interested in the following type
of polynomials.

\begin{definition}
We say that a polynomial
$$\pq(\pol{g},\pol{f}) := \pol{g}^{\deg{\pol{q}}}\cdot q\(\frac{\pol{f}}{\pol{g}}\)$$
is a \emph{candidate} if:
\begin{enumerate}
\item  $\pol{g}, \pol{q}, \pol{f}\in F[\lam]$,
\item $\pol{f}$ and $\pol{g}$ are relatively prime, and
\item $\pol{q}$ is monic and irreducible.
\end{enumerate}
\end{definition}

\forget
\note{Why do we need this lemma and the following proposition?}
\begin{lemma}\label{primeto}
If $\pol{Q} = \pq(\pol{g},\pol{f})$ is a candidate, then $\pol{Q}$ is prime to $\pol{g}$.
\end{lemma}
\begin{proof}
Write $\pol{q}= q_0 + q_1\lam + \dots + q_m\lam^m$. Then
$\pol{Q} = \sum{q_i \pol{f}^i \pol{g}^{m-i}} \equiv q_0 \pol{f}^m \pmod{\pol{g}}$.
As $\pol{q}$ is irreducible,  $q_0\neq 0$. But $\pol{f}$ is prime to $\pol{g}$.
\end{proof}

\begin{proposition}
Assume that $\pol{Q} = \pq(\pol{g},\pol{f})$ is a candidate. Then the degree of any factor of $Q(\lam)$ over $F$
is divisible by $\deg\pol{q}$.
\end{proposition}
\begin{proof}
Let $\pol{p}$ be an irreducible factor of $Q(\lam)$ over $F$,
let $K$ be the splitting field of $p$ over $F$,
and let $\gamma \in K$ be a root of $\pol{p}$.
By Lemma \ref{primeto}, $p$ is prime to $g$, therefore $g(\gamma)\neq 0$.
Let $\alpha = \frac{f(\gamma)}{g(\gamma)}$, and compute that
$g(\gamma)\cdot q(\alpha) = g(\gamma)\cdot q\(\frac{f(\gamma)}{g(\gamma)}\) = Q(f(\gamma),g(\gamma)) = 0$, thus $q(\alpha) = 0$.
Since $\pol{q}$ is irreducible (and $F$ is finite),
it follows that $q$ splits in $K$, and
$\deg\pol{q}$ divides $[K:F] = \deg\pol{p}$.
\end{proof}
\note{you still didn't answer the previous question}
\forgotten

\begin{theorem}
Assume that $\pol{Q} = \pq(\pol{g},\pol{f})$ is a candidate,
and let $\alpha$ be a root of $\pol{q}$ in the splitting field
$L$ of $\pol{q}$.
Then the number of distinct irreducible factors of $\pol{Q}$ over $F$ is
equal to the number of distinct irreducible factors of $\pol{f}-\alpha\pol{g}$ over $L$.
\end{theorem}
\begin{proof}
Denote by $\alpha_1,\cdots,\alpha_{m} \in L$ the (distinct) roots
of $\pol{q}$ in $L$, so that $\pol{q}$ has the form $\prod_{i=1}^{m}(\lam-\alpha_i)$.
We have that over $L$,
\begin{equation}\label{Qdec}
\pol{Q} = \pol{g}^m\cdot q\(\frac{\pol{f}}{\pol{g}}\) = \prod_{i=1}^{m}{\(\pol{f}-\alpha_i\pol{g}\)}.
\end{equation}

\smallskip

We can extend the standard norm map $L\to F$ to
a norm $N_{L/F}:L[\lam]\to F[\lam]$ by $N_{L/F}(\pol{h}) = \prod_{\s \in Gal(L/F)}{\s(\pol{h})}$,
where $\s(\lam)=\lam$ for all $\s\in Gal(L/F)$.
Fix any $\alpha \in \{\alpha_1,\dots,\alpha_m\}$.
Using this notation, Equation (\ref{Qdec}) is
$$Q(\lam) = N_{L/F}(\pol{f}-\alpha\pol{g}).$$
We will use the following lemma.

\begin{lemma}\label{norm}
Let the field $L$ be an extension of $F$.
Assume that $\pol{r} \in L[\lam]$ be irreducible.
Then $\pol{R} = N_{L/F}(\pol{r})$ is equal to an irreducible polynomial over $F$ raised to the power $[L:L_0]$,
where $L_0\sbst L$ is the subfield generated by the coefficients of $\pol{r}$ over $F$.
\end{lemma}
\begin{proof}
Since $N_{L/F} = N_{L_0/F}\circ N_{L/L_0}$ and $N_{L/L_0}(\pol{r}) = \pol{r}^{[L:L_0]}$, it is enough to
prove the claim in the case $L_0 = L$.

Let $\pol{R} = \pol{R_1} \cdots \pol{R_t}$ be an irreducible factorization of $\pol{R}$
over $F$. Obviously $\pol{r}$ divides $\pol{R}$ in $L[\lam]$, and
since $\pol{r}$ is irreducible we have that $\pol{r}$ divides one of
the factors, say $\pol{r}$ divides $\pol{R_1}$.

Let $L_1$ be the splitting field of $\pol{R_1}$ over $F$.
Note that $L \sbst L_1$, since the coefficients of $\pol{r}$ (which
divides $\pol{R_1}$) generate $L$. Let $L_2$ be the splitting
field of $\pol{r}$ over $L$, then $L_2 \sbst L_1$
and $\deg\pol{R} = [L:F]\cdot \deg\pol{r} = [L_2:F]$ divides
$[L_1:F] = \deg\pol{R_1}$.
Thus $\pol{R} = \pol{R_1}$ and is irreducible.
\end{proof}

Let $$\pol{f}-\alpha \pol{g} = \pol{u_1}^{s_1} \cdots \pol{u_t}^{s_t}$$ be
a factorization into irreducible polynomials over $L$.

Taking the norm from $L[\lam]$ to $F[\lam]$, we get the factorization
$$Q(\lam) = \pol{U_1}^{s_1} \cdots \pol{U_t}^{s_t}$$
over $F$, where $\pol{U_i} = N_{L/F}(\pol{u_i})$.
By Lemma \ref{norm}, The polynomials $\pol{U_i}$ are irreducible
(the coefficient of $\lam^{\deg\pol{g}}$ in $\pol{f}-\alpha\pol{g}$ generates $L$).
It thus remains to show that the $\pol{U_i}$ are relatively prime.
We will show that $u_i$ is prime to $\s(u_j)$ for any $\s \in Gal(L/F)$
and $j \neq i$. Indeed, if $\s = 1$ then $u_i$ is prime to
$u_j$ by the assumption. Otherwise, $u_{i}$ divides $f-\alpha g$ and
$\s(u_j)$ divides $f-\s(\alpha)g$, but $f-\alpha g$ and
$f-\s(\alpha)g$ are distinct and irreducible, thus relatively prime.
\end{proof}

%

\begin{corollary}\label{faster}
Assume that $\pol{Q} = \pq(\pol{g},\pol{f})$ is a candidate,
and let $L$ be the splitting field of $\pol{q}$.
Let $\alpha$ be a root of $\pol{q}$ in $L$. Then
$\pol{Q}$ is irreducible over $F$ if, and only if,  $\pol{f}-\alpha \pol{g}$ is irreducible over $L$.
\end{corollary}

According to \cite[Chapter 4]{ENC},
checking irreducibility of a degree $d$ polynomial
amounts to performing gauss elimination of a matrix of size
$d\x d$. In a finite field $F$ this requires roughly $d^3$
operations of multiplication and addition.
Assume that $\pol{Q} = \pq(\pol{g},\pol{f})$ is a candidate, and set $n = \max\{\deg\pol{f}, \deg\pol{g}\}$.
Checking the reducibility of $\pol{Q}$ directly over $F$ requires
roughly $\deg\pol{Q}^3 = m^3n^3$ operations. Checking its reducibility via
Corollary \ref{faster} requires roughly $n^3$ operations,
but here multiplication is more
expensive: each multiplication in $L$ requires roughly
$m^2$ multiplications in $F$.
Thus, the algorithm implied by Corollary \ref{faster} is roughly $m$ times faster, where $m=\deg\pol{q}$.
See also Remark \ref{wordparallelism}.

\forget
\section{Results of related interest}
If $\pol{f_S} = g(\lam ^t)$ for some $t$ dividing $n$,
then $\pol{f_{\<T,S\>}} = h(\lam ^t)$ for some $h$, thus cannot be primitive.
\note{details...}

What if we knew $\pol{q}$ is primitive?
\note{This is certainly not enough.}
\forgotten

\section{Primitivity}

Assume that $\pol{Q} = \pq(\pol{g},\pol{f})$ is a candidate,
$L$ is the splitting field of $\pol{q}$, and
$\alpha$ is a root of $\pol{q}$ in $L$.
By Corollary \ref{faster}, $\pol{Q}$ is irreducible over $F$ if, and only if,  $\pol{f}-\alpha \pol{g}$ is irreducible over $L$.
The analogue result for primitivity follows: $\pol{Q}$ is primitive if, and only if, it is irreducible and its roots generate $K^*$, where
$K$ is the splitting field of $\pol{Q}$.
Now, observe that $K$ is also the splitting field of $\pol{f}-\alpha\pol{g}$, and that $\pol{Q}$ and $\pol{f}-\alpha\pol{g}$
share the same roots in $K$.
This result, however, does not yield an improvement of the algorithm stated in the previous section.

In this section we show that if $f(0)=0$ and the base field is $F = \bits$
(these assumptions hold in the intended environment for the TSR),
then a candidate $Q(\lam) = \pq(\pol{g},\pol{f})$ is primitive only if $q$ is primitive.
Thus, the TSR-generation algorithm should begin with \emph{primitive} transformations $T$,
yielding an additional speedup factor $\phi(|L^*|)/|L^*|$, which is roughly $2$ when $\deg\pol{q}$
is a power of $2$, cf.\ \cite{ENC}.

It will be convenient to use the following definition.
\begin{definition}
Let $L$ be a finite field. The \emph{index} of a nonzero element $\alpha\in L$ is the index $|L^*|/|\<\alpha\>|$ of the cyclic group generated by $\alpha$ as a subgroup of
$L^*$.
\end{definition}
An irreducible polynomial is primitive if, and only if, its roots have index $1$ in its splitting field.
Note further that for $d$ dividing $|L^*|$, $\alpha\in L$ has index $d$ if, and only if, $\alpha = g^d$ for some generator $g$
of the cyclic group $L^*$.

\begin{lemma}\label{frobenius}
Let $h(\lam) \in L[\lam]$ be an irreducible monic polynomial of degree $n$
over $L$, with splitting field $K$ and a root $\mu$.
Then $\mu^{{|K^*|}/{|L^*|}} = (-1)^n h(0)$.
\end{lemma}
\begin{proof}
Let $\mu_0,\dots,\mu_{n-1}$ denote the (distinct) roots of $\pol{h}$.
Then $h(\lam) = (\lam-\mu_0)\cdots (\lam-\mu_{n-1})$ is the factorization over
$K$, thus $h(0) = (-1)^n \mu_0 \cdots \mu_{n-1}$.
On the other hand, the Galois group of $K/L$ is generated by the
Frobenius automorphism $u \mapsto u^{|L|}$, thus the roots of $\pol{h}$ are
$\mu,\mu^{|L|},\dots,\mu^{|L|^{n-1}}$, and
$\mu_0 \cdots \mu_{n-1} = \mu^{1+|L|+\dots+|L|^{n-1}} = \mu^{\frac{|L|^n-1}{|L|-1}}$.
\end{proof}

\begin{theorem}\label{overbits}
Assume that $F=\bits$ and $\pol{Q} = \pq(\pol{g},\pol{f})$ is an irreducible candidate
with $f(0)=0$.
If $\pol{q}$ is not primitive then $\pol{Q}$ is not primitive.
\end{theorem}
\begin{proof}
Let $K$ be the splitting field of $\pol{Q}$ over $F$, and
$L \sbst K$ the splitting field of $\pol{q}$.
Let $\mu \in K$ be a root of $\pol{Q}$, and $\alpha \in L$ a root
of $\pol{q}$.

Let $d_\mu$ denote the index of $\mu$ in $K$, and
$d_\alpha$ the index of $\alpha$ in $L$.
We will show that $d_\alpha = (|L^*|,d_\mu)$. Thus, $d_\mu = 1$ implies $d_\alpha = 1$.

By Corollary \ref{faster}, $\pol{h} = \pol{f}-\alpha \pol{g}$ is irreducible
over $L$.
Since every polynomial is monic over $\bits$, we can apply Lemma \ref{frobenius}
to get that
$h(0) = \mu^{{|K^*|}/{|L^*|}}$. But $h(0) = f(0)-(-1)^n \alpha g(0) = \alpha g(0)$.
As $\pol{f}$ and $\pol{g}$ are relatively prime, $g(0)\neq 0$, thus $g(0)=1$, and $h(0)=\alpha$.

Let $g$ be a generator of $K^*$ such that $\mu = g^{d_\mu}$.

Then $\alpha = \mu^{|K^*|/|L^*|} = g^{d_\mu|K^*|/|L^*|}$,
and its order in $K^*$ is
$$|K^*|/(|K^*|,d_\mu|K^*|/|L^*|) = |L^*|/(|L^*|,d_\mu)\mbox{\rm ,}$$
as asserted.
\end{proof}

\section{The final generation algorithm}

In light of the results obtained in the previous sections,
we end up with the following algorithm for TSR-generation over $F=\bits$:
\begin{algorithm}[Primitive TSR generation]\label{algo}
\
\begin{enumerate}
\item Choose at random a primitive transformation $T$ on $\bits^m$.
\item Choose a random sequence $S=\<a_0,\dots,a_{n-1}\>\in \bits^n$ such that $a_{0}\neq 0$.
\item Choose a root $\alpha$ of $\pol{f_T}$ in its splitting field $L$.
\item Check that $\lam^n-\alpha \pol{f_S}$ is irreducible over $L$ (otherwise return to step 1).
\item Check that $\pol{Q} = \p_{\pol{f_T}}(\pol{f_S},\lam^n)$ is primitive: Choose a root $\mu$
of $\pol{Q}$ in its splitting field $K$, and check for all prime $p$ dividing $|K^*|$ that $\mu^{|K^*|/p}\neq 1$
(in fact, as we show below, it is not needed to consider the cases where $p$ divides $|L^*|$).
\item If $\pol{Q}$ is not primitive, return to step 1.
\end{enumerate}
\end{algorithm}

\begin{remark}\label{repeat}
Assuming that generally, the probability that $\pol{Q} = \p_{\pol{f_T}}(\pol{f_S},\lam^n)$ is primitive
is roughly the same for every primitive transformation $T$, it would be more efficient to repeat steps 2 to 5 of the algorithm
several times before starting again from step 1. Thus, the complexity of step 1 will be negligible with respect to the total
running time. Moreover, we argue below that step 5 usually occurs only once.
\end{remark}

\begin{remark}\label{wordparallelism}
In all of the mentioned algorithms, one can get a
speedup factor of $\tilde m$, where $\tilde m$ is the size of the word in the processor
where the search for the TSR is made (note that this need not be the same processor on which the TSR will be
implemented, thus $\tilde m$ need not be equal to $m$). This is done by exploiting the processors word-oriented
operations to define parallel versions of the basic operations used in the algorithms.
\end{remark}

For a natural number $n$, we denote by $C_n$ the (multiplicative) cyclic group of order $n$.
If $g$ is a generator of $C_n$, then $g^x$ is a generator as well if, and
only if, $(x,n)=1$. This is why the number of generators of $C_n$ is exactly
$\phi(n)$, where $\phi$ is Euler's function, and the probability that a uniformly chosen
element generates $C_n$ is $\phi(n)/n$.
An irreducible polynomial $\pol{Q}$ is primitive if a root $\mu$ of $\pol{Q}$
generates the multiplicative group of its splitting field $K$.
There is a natural $1$ to $[K:F]$ correspondence between irreducible monic polynomials of
degree $[K:F]$ and elements of $K$ which do not belong to a proper subfield
of $K$. This correspondence implies that the probability that an irreducible $\pol{Q}$ is primitive is
close to $\phi(|K^*|)/|K^*|$.

We now consider \emph{irreducible candidates}. We wish to estimate the probability that
a candidate passing the test in step 4 of the algorithm will also past the final test of step 5.
A candidate $\pol{Q} = \p_{\pol{f_T}}(\pol{f_S},\lam^n)$ is \emph{good} if
$T$ is primitive and $\pol{Q}$ is irreducible. We will find, heuristically, the
probability that a good candidate is primitive.
Let $L$ be the splitting field of $\pol{f_T}$, and $K$ be the splitting field of $\pol{Q}$.
Factor $|K^{*}| = k_L \cdot a$, where $k_L$ is the product of all the prime factors of $|K^{*}|$ which divide $|L^{*}|$
(allowing powers of primes).
Then the group $K^*$ is isomorphic to $C_{k_L} \times C_a$, where a prime $p$ divides $|L^*|$ if, and only if, it divides $k_L$.
A root $\mu$ of $\pol{Q}$ generates $K^*$ if, and only if, its projections in $C_{k_L}$ and $C_a$ are both
generators.

In the proof of Theorem \ref{overbits} we showed that $d_\alpha$, the co-order of a root $\alpha$ of $\pol{f_T}$ in $L$,
is equal to $(|L^*|,d_\mu)$, where $d_\mu$ is the co-order of $\mu$ in $K$.
As $T$ is primitive (i.e.\ $d_\alpha = 1$), we have that $d_\mu$ is prime to $|L^*|$.
Thus, $d_\mu$ is prime to $k_L$, that is, $k_L$ divides the order of $\mu$ in $K^*$.
Therefore, the projection of $\mu$ in $C_{k_L}$ is a generator of that group.
We assume, \emph{herusitically},
that the projection of $\mu$ on $C_a$ is (close to being) uniformly distributed.
Thus, the probability of its being a generator of $C_a$ is close to
$\phi(a)/a$. In general,
$$\frac{\phi(n)}{n} = \prod_{p|n}{\(1-\frac{1}{p}\)}\mbox{\rm ,}$$
and as a prime $p$ divides $a$ if, and only if, $p$ divides $|K^*|$ but not $|L^*|$, we have that
$$\frac{\phi(a)}{a} = \frac{\phi(|K^*|)/|K^*|}{\phi(|L^*|)/|L^*|}.$$

We thus have a heuristic justification for the following claim.
\begin{claim}\label{highprob}
Assume that $\pol{Q} = \p_{\pol{f_T}}(\pol{f_S},\lam^n)$ is an irreducible candidate over $\bits$,
where $\pol{f_T}$ is primitive.
Then the probability that $\pol{Q}$ is primitive is close to
$$\frac{\phi(|K^*|)/|K^*|}{\phi(|L^*|)/|L^*|}.$$
\end{claim}

\begin{example}
The probability at Claim \ref{highprob} is usually close to 1. We give here a few examples:
\begin{enumerate}
\item When the word's size is $8$ bits and the number of words is $7$, we have that
$\phi(2^{56}-1)/(2^{56}-1)\approx 0.465$, $\phi(2^8-1)/(2^8-1)\approx 0.502$, and
the division yields probability close to $0.927$.
\item When the word's size is $16$ bits and the number of words is $4$, we get
probability close to $0.998$.
\item For values $24$ and $3$, respectively, we get $0.898$.
\item For values $32$ and $2$ we get $0.998$.
\end{enumerate}
\end{example}


\section{Concluding remarks}\label{conclusions}

We have presented the family of linear transformation shift registers which is efficient in
software implementations. The theory we developed enabled us to get an efficient algorithm
for generation of primitive transformations of this type (i.e., which have maximal period),
thus answering a challenge raised in \cite{FSE94}.

Variants of our construction can be found more appropriate
for certain applications.
Arguments similar to the ones we have presented here may be found useful
in the study of these variants as well.
A noteworthy variant of the LFSR type that we have studied is the
\emph{internal-xor}, or \emph{Galois}, shift register (See, e.g., \cite{SCH}).
The number of new bits generated in one step of
an internal-xor shift register is equal on average
to half of the number of taps in that LFSR.
Our construction suggests an obvious
analogue internal-xor TSR. We get exactly the same results
for this case,
since the characteristic polynomial of an internal-xor TSR
is equal to that of the corresponding external-xor TSR, which
we have studied in this paper.


\end{document}